\documentclass{aa}
\usepackage{color,graphicx}
\usepackage{txfonts}

\begin{document}

\title{Comparison of acoustic travel-time measurement of solar meridional circulation from SDO/HMI and SOHO/MDI}
\titlerunning{Comparison between MDI and HMI}

\author{
Zhi-Chao Liang \inst{\ref{mps}} \and
Aaron C. Birch \inst{\ref{mps}} \and
Thomas L. Duvall, Jr. \inst{\ref{mps}} \and
Laurent Gizon \inst{\ref{mps},\ref{gottingen}} \and
Jesper Schou \inst{\ref{mps}}
}

\institute{
Max-Planck-Institut f\"ur Sonnensystemforschung, Justus-von-Liebig-Weg 3, 37077 G\"ottingen, Germany\\
\email{zhichao@mps.mpg.de} \label{mps}\and
Institut f\"ur Astrophysik, Georg-August-Universit\"at G\"ottingen, Friedrich-Hund-Platz 1, 37077 G\"ottingen, Germany \label{gottingen}
}

\date{Received $\langle$date$\rangle$ / Accepted $\langle$date$\rangle$}

\abstract
{
Time-distance helioseismology is one of the primary tools for studying the solar meridional circulation, especially in the lower convection zone.
However, travel-time measurements of the subsurface meridional flow suffer from a variety of systematic errors, such as a center-to-limb variation and an offset due to the position angle ($P$-angle) uncertainty of solar images.
It has been suggested that the center-to-limb variation can be removed by subtracting east-west from south-north travel-time measurements.
This ad~hoc method for the removal of the center-to-limb effect has been adopted widely but not tested for travel distances corresponding to the lower convection zone.
} {
We explore the effects of two major sources of the systematic errors, the $P$-angle error arising from the instrumental misalignment and the center-to-limb variation, on the acoustic travel-time measurements in the south-north direction.
} {
We apply the time-distance technique to contemporaneous medium-degree Dopplergrams produced by SOHO/MDI and SDO/HMI to obtain the travel-time difference caused by meridional circulation throughout the solar convection zone.
The $P$-angle offset in MDI images is measured by cross-correlating MDI and HMI images.
The travel-time measurements in the south-north and east-west directions are averaged over the same observation period (May 2010 to Apr 2011) for the two data sets and then compared to examine the consistency of MDI and HMI travel times after applying the above-mentioned corrections.
} {
The offsets in the south-north travel-time difference from MDI data induced by the $P$-angle error gradually diminish with increasing travel distance.
However, these offsets become noisy for travel distances corresponding to waves that reach the base of the convection zone.
This suggests that a careful treatment of the $P$-angle problem is required when studying a deep meridional flow.
After correcting the $P$-angle and the removal of the center-to-limb effect, the travel-time measurements from MDI and HMI are consistent within the error bars for meridional circulation covering the entire convection zone.
The fluctuations observed in both data sets are highly correlated and thus indicate their solar origin rather than an instrumental origin.
Although our results demonstrate that the ad~hoc correction is capable of reducing the wide discrepancy in the travel-time measurements from MDI and HMI, we cannot exclude the possibility that there exist other systematic effects acting on the two data sets in the same way.
} {}


\maketitle

\section{Introduction}

The meridional circulation is a large-scale flow in the meridional plane of the Sun that overall has an antisymmetric profile about the equator.
Surface meridional flow can be determined by direct Doppler shift measurements \citep[e.g.,][]{Duvall1979,Hathaway1996a,Ulrich2010} or by tracking photospheric features \citep[e.g.,][]{Komm1993,Hathaway2010}.
The surface meridional flow is poleward with a peak speed of 10--20~m~s$^{-1}$ at midlatitudes.
The subsurface meridional circulation can be measured with helioseismic methods, for example, the Fourier-Legendre method \citep{Braun1998}, mode eigenfunction perturbation analysis \citep{Woodard2013}, ring-diagram analysis \citep{Haber2002}, and time-distance analysis \citep{Giles1997}.
However, it has been a challenging task to obtain a precise measurement of the deep meridional flow because of the small signal-to-noise level \citep{Braun2008} and a plethora of systematic errors \citep[e.g.,][]{Gizon2003,Duvall2009,Larson2015}.
Near the surface, local helioseismic measurements of meridional circulation generally agree with the surface measurements mentioned above.
However, the various helioseismic methods disagree about the subsurface structure of the meridional circulation \citep{Zhao2013,Schad2013,Jackiewicz2015,Rajaguru2015}.
Since the large-scale meridional circulation plays a critical role in flux-transport dynamo models, resolving this issue is a priority \citep[e.g.,][]{Charbonneau2014,Cameron2016}.
An additional complication is introduced by the variation of the meridional circulation with solar activity levels \citep[e.g.,][]{Chou2001,Beck2002,Gizon2004a,Zhao2004}.

The space-based instrument Michelson Doppler Imager on board the \textit{Solar and Heliospheric Observatory} \citep[SOHO/MDI,][]{Scherrer1995} and its successor the Helioseismic and Magnetic Imager on board the \textit{Solar Dynamical Observatory} \citep[SDO/HMI:][]{Scherrer2012,Schou2012} have collected nearly uninterrupted observations of the Sun over two decades since 1996.
While MDI recorded the Doppler shift of the \ion{Ni}{i} 6768~$\AA$ spectral line at a cadence of 60 seconds, HMI measures that of the \ion{Fe}{i} 6173~$\AA$ line at a 45-second cadence.
Numerical simulations suggest that HMI Doppler velocity signal is formed at 150--200~km when taking into account the limited spatial resolution of the instrument \citep{Fleck2011,Nagashima2014} and the MDI Doppler velocity signal is formed about 25~km higher than HMI \citep{Fleck2011}.
There is an overlap of one year between the two missions.
These contemporaneous data observed at different spectral lines offer an excellent opportunity to assess the reliability of the measurement of the deep meridional flow and facilitate examining the systematic errors in time-distance helioseismology.

It has been reported that there is an error in MDI position angle ($P$-angle) due to instrumental misalignment \citep{Giles1997,Giles2000,Beck2005}.
This $P$-angle error causes a leakage of the solar rotation signal into the meridional flow measurement and produces a 0.5-second offset in south-north travel-time differences for near-surface measurements.
An early estimate of the MDI $P$-angle error was obtained by \citet{Evans1999} who compared MDI images with observations from the Mount Wilson Solar Observatory and found that MDI images are rolled by $0.32\degr\pm0.05\degr$ with respect to the Mount Wilson images.
Later on, smaller values, $0.19\degr\pm0.04\degr$ and $0.22\degr\pm0.03\degr$, were obtained by cross-correlating MDI and GONG images using observations from May 2000 and the Mercury transit in November 1999, respectively\footnote{
MDI Calibration Notes and Known Problems (roll angle error from GONG comparisons; 2000), http://soi.stanford.edu/data/cal}.
\citet{Hathaway2010} took a different approach and deduced a $P$-angle offset of 0.21$\degr$ from the cross-equatorial signal of the surface meridional flow measured by tracking photospheric features.
Recently, \citet{Schuck2016} compared MDI magnetograms with those from HMI during June 2010 and reported a time-dependent offset between the two cameras varying from 0.18$\degr$ to 0.23$\degr$.
Since a small $P$-angle misalignment could yield a noticeable offset in the south-north travel-time measurement, a $P$-angle correction has to be carried out when comparing results of meridional flow from different instruments.

Another major source of systematic errors in time-distance analysis is the center-to-limb effect, which is an additional variation that is a function of the center-to-limb angle superimposed on the travel-time differences.
\citet{Duvall2009} extended the study of \citet{Giles2000} of time-distance measurements of deep meridional flow but stumbled across a center-to-limb variation instead.
\citet{Zhao2012} reported that different observables (with different formation heights in the solar atmosphere) have different center-to-limb variations.
Since the solar rotation rate does not change along the equator, the east-west travel-time measurement with longitude would be expected to capture the center-to-limb variation.
Thus, they adopted an ad~hoc method to remove the center-to-limb effect by subtracting east-west measurements from south-north measurements.
In that work, this correction method was examined for travel distances less than 3.84$\degr$ (heliocentric degree) and brought the diverse measurements of meridional flow using different observables to a remarkable consistency.
\citet{Zhao2013} further applied the empirical method to travel-time measurements from MDI and HMI data but not in a concurrent period.
They found that the travel-time measurements from MDI data are roughly comparable to those from HMI for travel distances ranging from 2$\degr$ to 38$\degr$ after the removal of the center-to-limb effect.
The cause of the center-to-limb variation is still unclear, but might perhaps be explained, to some extent, by the asymmetry between the upflows and downflows in the near-surface granular convection \citep{Baldner2012}.
Considering the important role of center-to-limb variations in studying the deep meridional flow, it is necessary to scrutinize this empirical correction method for waves that probe the entire convection zone.

In this study, we apply the time-distance technique to contemporaneous Dopplergrams taken by MDI and HMI to examine the aforementioned systematic errors.
In Sect.~\ref{sec:data}, we describe the data preparation, measurement of the $P$-angle offset, and travel-time measurement.
The influences of the $P$-angle error and the center-to-limb variation on travel-time differences are presented in Sect.~\ref{sec:result}, and the implications of the results are discussed in Sect.~\ref{sec:discuss}.

\section{Data and analysis}
\label{sec:data}

We analyze full-disk Dopplergrams from the MDI and HMI medium-$\ell$ data in the 1-yr period from 1 May 2010 to 30 April 2011, which is the overlap interval between the two missions.
The MDI images are retrieved from the \textsf{mdi.vw\_v} data series in the Joint Science Operations Center (JSOC) data system while HMI images are from the \textsf{hmi.vw\_v\_45s} data series.
The \textsf{mdi.vw\_v} data are not corrected for image distortion \citep[see][]{Larson2015} nor are they corrected for modulation transfer function (MTF) variation across the field of view \citep[see][]{Tarbell1997,Korzennik2004}.
We select dates with a duty cycle of more than 75\%.
In addition, if changes of keywords \texttt{CDELTn} and \texttt{CROTA2} in a single day are greater than $5\times10^{-3}$~arcsec~pixel$^{-1}$ and 0.3$\degr$, respectively\footnote{
The two threshold values roughly correspond to half a pixel at the edge of the solar images and help pick out two days of data due to changes in MDI focus position and five days of data when either of the instruments was rotating.
}, that date will be discarded.
After the preliminary reduction including the Fourier filtering (described below), we compute the mean value and standard deviation of pixels within a 90\% radius of the solar image for each frame.
Any day in which these numbers exceed empirical thresholds (5~m~s$^{-1}$ for the mean and 70~m~s$^{-1}$ for the standard deviation) is investigated by hand to further reject suspicious data\footnote{
This step picks out two days of data due to MDI tuning changes.
In future work the steps of checking keywords and statistical numbers will be replaced by examining the event tables of the instruments directly, which would be a more straightforward method to exclude the bad data.
}.
Because a number of MDI images are missing or of poor quality in the last few months of its mission, only 288-day contemporaneous data are selected for this study.

The medium-$\ell$ images from both data sets have an image scale of $\sim$10~arcsec~pixel$^{-1}$ or $\sim$0.6~heliographic degree pixel$^{-1}$ at the disk center.
These images result from Gaussian-weighted binning of the full-resolution Dopplergrams and are sensitive to solar $p$-modes up to harmonic degree $\ell=300$ \citep{Kosovichev1997}.
Most of the background patterns in the Dopplergrams, such as the solar rotation and the velocity of the observer, are removed by subtracting a simple moving average of 1~hr for each pixel.
A high-pass temporal filter tapered by a half-cycle raised-cosine function between 1.5 and 2~mHz is further applied to remove unwanted signals such as granulation and supergranulation.
Most of the $f$ modes are removed as well because of this high-pass filter.
Also, a low-pass filter tapered between 6.5 and 7~mHz is used because of a low-$\ell$ artifact at $\sim$7.4~mHz in HMI Dopplergrams.
No Fourier-filter in the spatio-temporal domain (e.g., phase-speed filter) is applied in this study to avoid the spreading of the magnetic field, where a masking procedure is performed in the following analyses to exclude active regions.

\begin{figure}
\centering
\resizebox{\hsize}{!}{\includegraphics{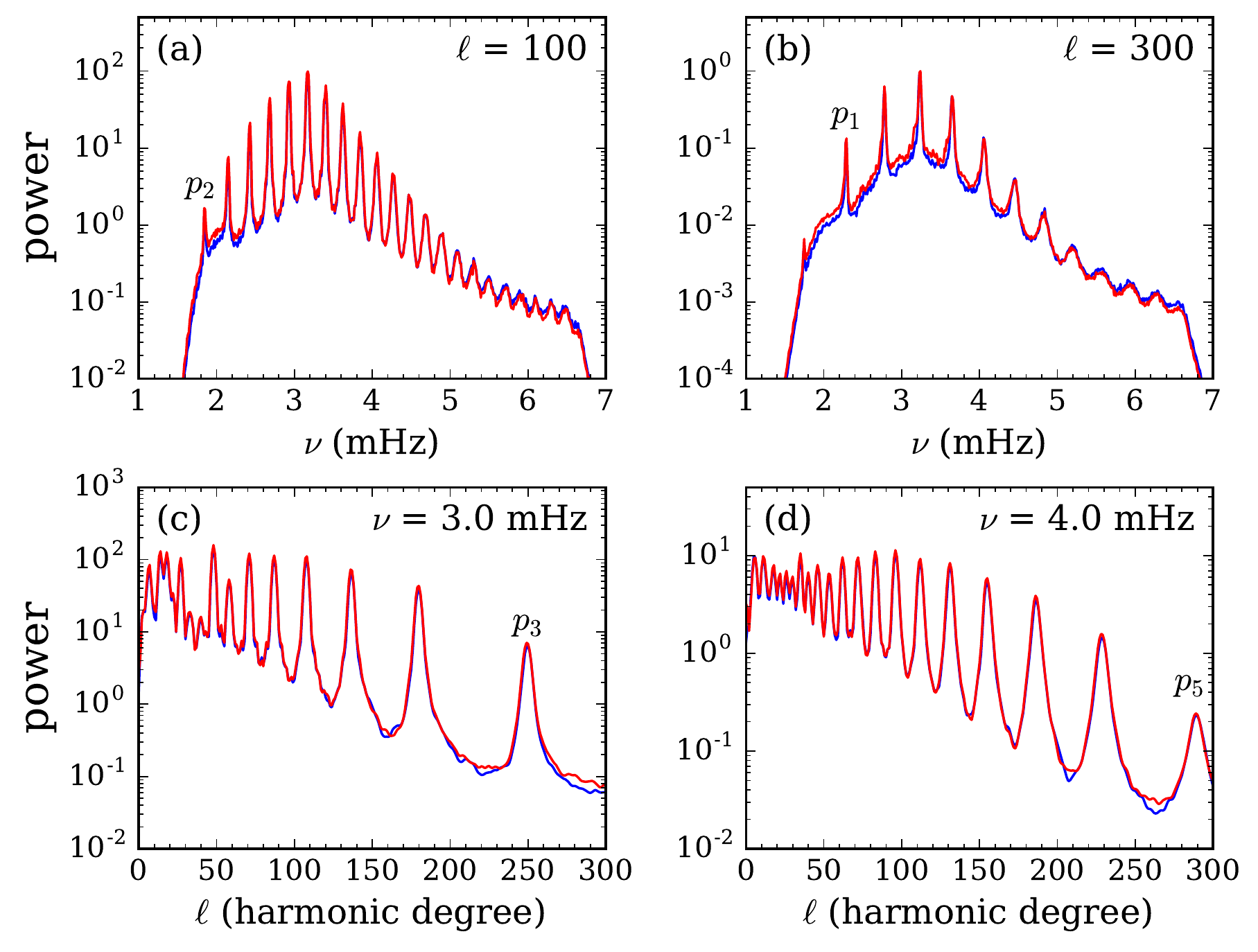}}
\caption{ \label{fig:spect}
Comparisons of $m$-averaged power spectra between MDI (blue) and HMI (red) obtained from one-day tracked Dopplergrams.
The upper panels show the power as a function of frequency at selected harmonic degrees: (a) $\ell=100$ and (b) $\ell=300$.
The lower panels show the power as a function of harmonic degree at selected frequencies: (c) 3.0~mHz and (d) 4.0~mHz.
Selected modes have been labeled.
The power is normalized by the square of the number of frames.
}
\end{figure}

Fig.~\ref{fig:spect} shows comparisons of $m$-averaged power spectra between MDI and HMI for selected frequencies and harmonic degrees.
The power spectra are computed from spherical harmonic decomposition of one-day tracked Dopplergrams (the image tracking is described in Sect.~\ref{trk}).
Although the noise level of HMI is slightly higher than that of MDI for $\ell > 200$, the profiles from both data sets are generally in line with each other.
The difference between the two profiles could be due to different formation heights and thus different sensitivities to the $p$-mode oscillations.
The MTFs of the two instruments may also contribute to the deviation in their spectral amplitudes.
The differences between the power spectra of the HMI and MDI Doppler observations, combined with (potential) $\ell$ and frequency dependence of the phase-shift between Doppler observations from HMI and MDI, may play a role in causing the difference in the center-to-limb variations in travel times measured from MDI and HMI observations.
A detailed investigation of the interplay between formation heights, $p$-mode spectra, and the center-to-limb variations is beyond the scope of this article.

\subsection{$P$-angle offset} \label{roll}
The $P$-angle of HMI data is known to be accurate to a few thousandths of a degree by virtue of the Venus transit of 5--6 June 2012 \citep{Couvidat2016}.
A further analysis of HMI data of the Mercury transit on 9 May 2016 using a similar method shows that the roll difference between the results of the two events is less than $0.003\degr$, and we therefore assume that the roll angle of the HMI instrument is correct.
We then determine the $P$-angle offset of MDI by comparing MDI images to contemporaneous HMI images.
Doppler signals of solar $p$-mode oscillations obtained from the preliminary analysis are favored over magnetograms\footnote{
In fact we have also tried using the MDI 96-minute magnetograms compared with HMI magnetograms but the fluctuations in the measured offsets are somewhat larger than using the MDI 60-second Dopplergrams compared with HMI Dopplergrams; the MDI 60-second magnetograms only have a duty cycle less than 10\% in this period so we try using the 96-minute magnetograms instead.}
because different formation heights might change the character of derived magnetic fields \citep{Evans1999} and, in addition, the correlation depends on the field strength in the analyzed areas \citep{Liu2012}.
We add a set of incremental angles to the nominal $P$-angle of MDI with a step size of 0.01$\degr$ over the range 0.1$\degr$--0.3$\degr$ and remap the $P$-angle adjusted MDI images to have the same Carrington longitude, solar tilt angle $B_0$, nominal $P$-angle, and spatial resolution as that of HMI at the disk center.
The Pearson correlation coefficients are then computed between the remapped MDI and HMI images using pixels within the center-to-limb angle of 60 heliocentric degrees.
The maximum value of the correlation coefficients and the offset in $P$-angle producing the strongest correlation are measured by fitting a parabola through the three closest points to the correlation coefficient peak.
The above procedures are repeated for the 1-yr long data at 30-minute intervals.

\begin{figure}
\centering
\resizebox{\hsize}{!}{\includegraphics{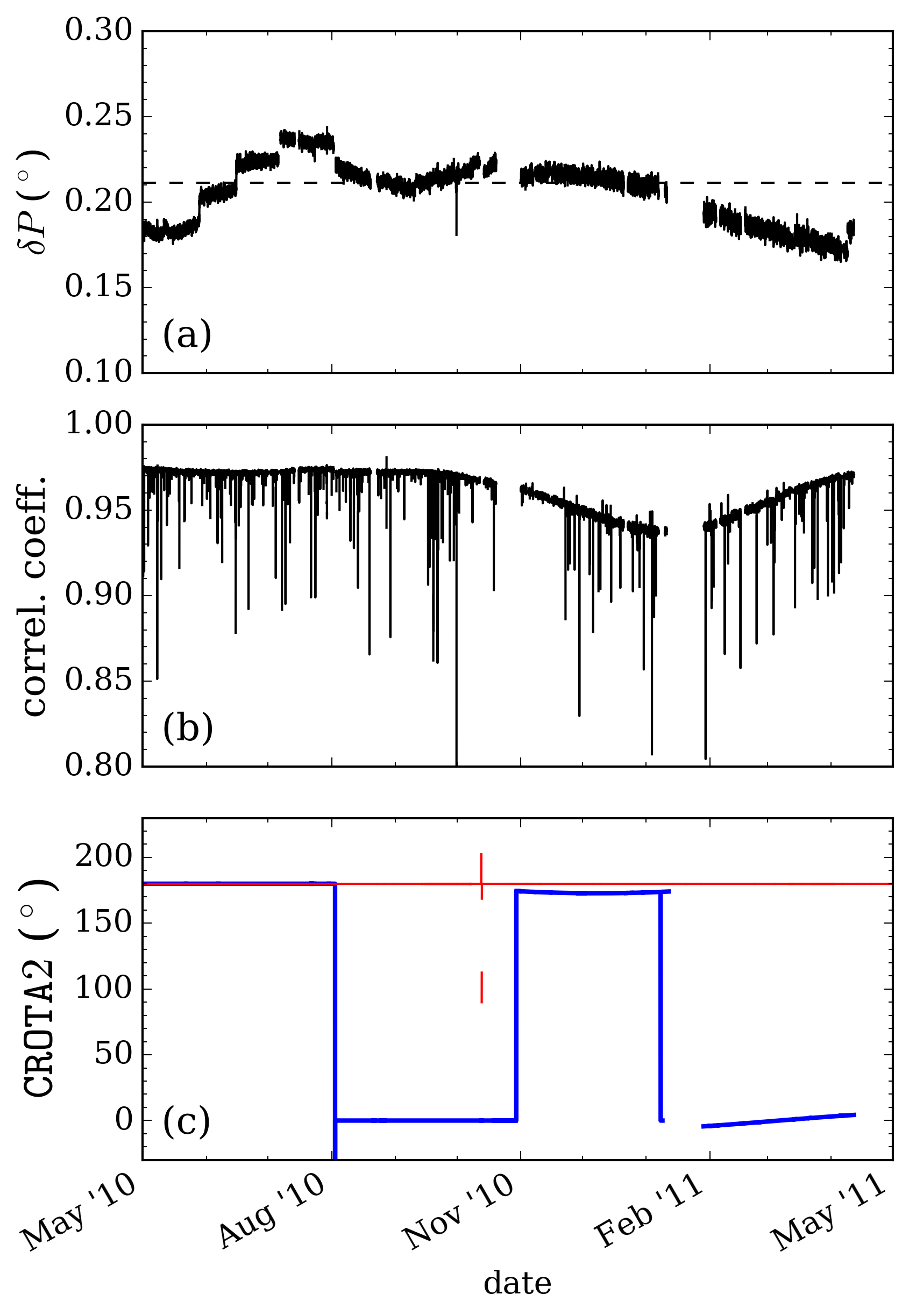}}
\caption{ \label{fig:reg}
(a) $P$-angle offset $\delta P$ of MDI with respect to HMI (front camera) as a function of time.
The horizontal dashed line indicates the median value of $\delta P$ in this period.
(b) Maximum values of Pearson correlation coefficients corresponding to the measurements in panel (a).
(c) The keyword \texttt{CROTA2} (nominal $P$-angle with sign reversed) for MDI (blue) and HMI (red) in the meantime.
}
\end{figure}

Fig.~\ref{fig:reg}a and \ref{fig:reg}b show the MDI $P$-angle offset $\delta P$ with respect to HMI (front camera) and the maximum value of the Pearson correlation coefficients determining the $\delta P$ as a function of time, respectively.
The keyword \texttt{CROTA2} (nominal $P$-angle with sign reversed by its definition) from the MDI and HMI data records in the corresponding period are also plotted in Fig.~\ref{fig:reg}c.
The MDI instrument was nominally aligned with the solar rotation axis since the deployment of SOHO spacecraft in 1996 and started to flip 180$\degr$ every three months as of mid-2003.
Accordingly, MDI's \texttt{CROTA2} is a fixed value of 0$\degr$ or 180$\degr$ alternately after mid-2003.
But the flight operation of SOHO spacecraft was changed so that the spacecraft pointed to the ecliptic pole starting 29 October 2010\footnote{
SOHO Ancillary Data: Attitude, http://sohowww.nascom.nasa.gov/data/ancillary/\#attitude}.
This implies that MDI's \texttt{CROTA2} is no longer a fixed value after this point.
The measured $\delta P$ gradually drifts between 0.18$\degr$ and 0.24$\degr$ for the most part in this period.
Clearly, some of the sudden changes in $\delta P$ coincide with the alteration of the SOHO spacecraft operation, though not all of the discontinuities in Fig.~\ref{fig:reg}a are concurrent with those in Fig.~\ref{fig:reg}c.
Also, the discontinuity of $\delta P$ jumping from 0.21$\degr$ to 0.23$\degr$ in the middle of June appeared in the one month's measurement of the relative roll angle between MDI and HMI (side camera) magnetograms performed by \citet[, Fig.~1]{Schuck2016}.
The standard deviation of $\delta P$ in an individual day is $\sim$0.002$\degr$ but rises to $\sim$0.003$\degr$ in the last few months.
Also, the correlation coefficients start drifting when \texttt{CROTA2} deviates from constant values.

We have made an attempt to calibrate the notorious image distortion of the MDI instrument to maximize the correlation coefficients.
The correction for the elliptic distortion is performed as in \citet[, appendix]{Korzennik2004}, where the required parameters are the CCD tilted angle $\alpha$, the rotation angle $\beta$ describing the direction around which the CCD is tilted, and the effective focal length $f_\mathrm{eff}$.
With the values in their work ($\alpha=2.59\degr$ and $\beta=56\degr$), the calibration greatly enhances the correlation coefficients for the periods when SOHO was flipped but lowers the coefficients for alternate periods instead.
As a compromise, we adopt the values $\alpha=1.8\degr$ and $\beta=50\degr$ to improve correlations without causing notable harm for non-flipping periods, but the correlations are not as good as the results with their values for flipping periods.
The value of $f_\mathrm{eff}$ and the calibration parameters for cubic distortion are taken from a recent study by \citet{Larson2015}.
The calibration of MDI image distortion applied here helps achieve a better correlation and smaller fluctuations in the measured $\delta P$.
However, it does not account for the discontinuities and long-term variation of $\delta P$.

In this work we only apply a constant $P$-angle correction of 0.21$\degr$ (the median value of $\delta P$ in this period) to MDI images.
The residuals of time-varying $\delta P$ are relatively small and have little effect on our final results.

\subsection{Acoustic travel time measurement}
The full-disk Dopplergrams of solar $p$-mode oscillations derived from the preliminary reduction are mapped onto heliographic coordinates using equidistant cylindrical projection where both the heliographic longitude and latitude are evenly spaced with a map scale of 0.6$\degr$~pixel$^{-1}$.
The interpolation method used for the mapping is the bicubic spline \citep[e.g.,][, \S~3.6]{Press1992}.
For MDI images, an additional $P$-angle correction is applied while remapping.
Along with the mapping procedure, every 24-hr time series of images is tracked relative to the midday to remove the solar differential rotation at the surface \citep{Chou1999}.\label{trk}
The surface rotation rate adopted for each latitude $\lambda$ is $451.5-65.3\sin^2\lambda-66.7\sin^4\lambda$~nHz \citep[, Table~2, derived from \citet{Ulrich1988}]{Schou1998}.
Each resulting data cube has the size of $120\degr\times120\degr\times\,$24hr.

\begin{figure}
\centering
\resizebox{\hsize}{!}{\includegraphics{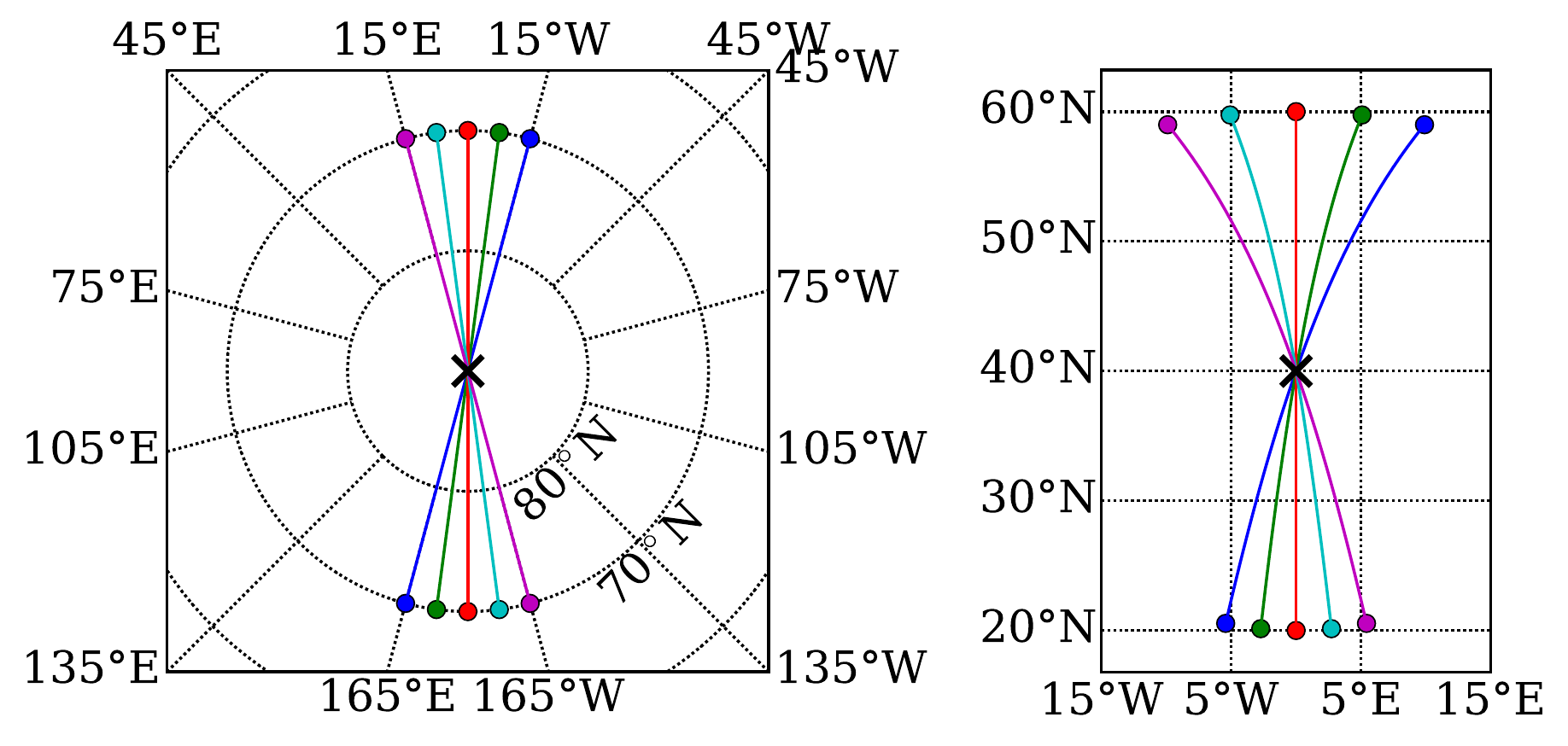}}
\caption{ \label{fig:arc}
\emph{Left}: Schematic plot of an arc-to-arc geometry in the polar projection with a travel distance of 40$\degr$ and an arc size of 30$\degr$ as an example.
The end points (solid points) on the two arcs all have the same distance of 20$\degr$ from the central point (cross symbol).
The solid lines connecting pairs of points show the great circles.
\emph{Right}: Same configuration as in the left panel but rotating the coordinates to a different latitude and transforming into cylindrical projection.
Whereas the distances and azimuths between points are preserved during the coordinate transformation, the arcs in the right panel are stretched due to the distortion introduced by the projection.
The cross-correlation functions between pairs of end points connected by the great circles are computed, averaged over different pairs of points, and then assigned to the central point.
The number of points on the arc shown here is less than in real cases for clarity.
}
\end{figure}

To probe a subsurface meridional flow, we utilize a common-midpoint deep-focusing scheme \citep{Duvall2003}, in which the cross-correlation function (CCF) between pairs of points is computed in an arc-to-arc geometry to improve the signal-to-noise ratio.
In order to determine the correct separation distance $\Delta$ between pairs of points, two arcs in the opposite direction with a subtended angle of 30$\degr$ and a distance of $\Delta/2$ from the central point are first laid in the polar projection as shown in the left panel of Fig.~\ref{fig:arc}.
This ensures that the distances and azimuths between these points are correct since all the lines passing through the center in this projection are great circles.
The arc length of the interval between adjacent points is set to be $\sim$0.6$\degr$ in polar projection.
The coordinates are then rotated and transformed into equidistant cylindrical projection such that the central point is at the desired location and the two arcs are in the south-north direction (or in the east-west direction for measurements of center-to-limb variations) as shown in the right panel of Fig.~\ref{fig:arc}.
For points not on the grid pixels, cubic convolution interpolation \citep{Keys1981} is used instead of bicubic spline to reduce the computational burden.
After acquiring the desired points in each frame, the CCFs between pairs of opposite points on the two arcs are computed, averaged, and then associated with the central points.
The above procedure is repeated for different central points with a step size of 0.6$\degr$ in longitude and latitude, and for different travel distances $\Delta$ covering 6$\degr$--50.4$\degr$ in 0.6$\degr$ increments.
The arc-averaged CCFs are further averaged over $\pm15\degr$~longitude for the south-north case or $\pm15\degr$~latitude for the east-west case, and then averaged over 288~days.

If any of the paired points is within an active region, the CCF of this pair is excluded in the average to avoid the influence of magnetic fields on the travel-time measurements \citep{Liang2015a}.
In practice if the field strengths within an area of $4\times4$~pixels nearest to the location of interest are greater than a certain threshold value (discussed below), the related CCF is discarded.
The daily magnetograms for determining the active regions are derived in much the same way as in \citet{Liang2015a} but with a $P$-angle correction for MDI data.
A threshold of 50~G is also used to identify the active regions for MDI, but a smaller threshold of 35~G is used for HMI since the magnetic signal of MDI data is greater than that derived from HMI data by a factor of $\sim$1.4 \citep{Liu2012}.
The difference in the number of masked pixels between MDI and HMI is less than 10\%.
\citet{Liang2015a} showed that the difference between the travel-time measurements with a threshold of 35~G and 50~G is less than 0.04~seconds during the solar maximum.
Therefore the masking procedure used here is not expected to be a major source of systematic error.

The northward and southward (or eastward and westward) phase travel times are determined by fitting a Gabor wavelet to a 20-minute interval in the vicinity of the first skip in 288-day averaged CCF \citep{Kosovichev1997a,Duvall1997}.
More precisely, the group travel time computed from the ray approximation is used as the reference time to determine the fitting window, and the position of CCF's peak closest to it is used as the initial guess for the phase time in the nonlinear least-squares fitting.
Details of the fitting window position are provided in Appendix~\ref{suppl:win}.
The differences between southward and northward phase travel times $\delta\tau_\mathrm{SN}$ (or differences between eastward and westward travel times $\delta\tau_\mathrm{EW}$) are then taken for each $\Delta$ and latitude (or longitude).
Here the subscript notations of $\delta\tau$, SN and EW, indicate the convention of southward minus northward travel times and eastward minus westward travel times, respectively.
Therefore a positive value of $\delta\tau_\mathrm{SN}$ in general corresponds to a northward flow for a near-surface measurement.

\section{Results}
\label{sec:result}

\begin{figure}
\centering
\resizebox{\hsize}{!}{\includegraphics{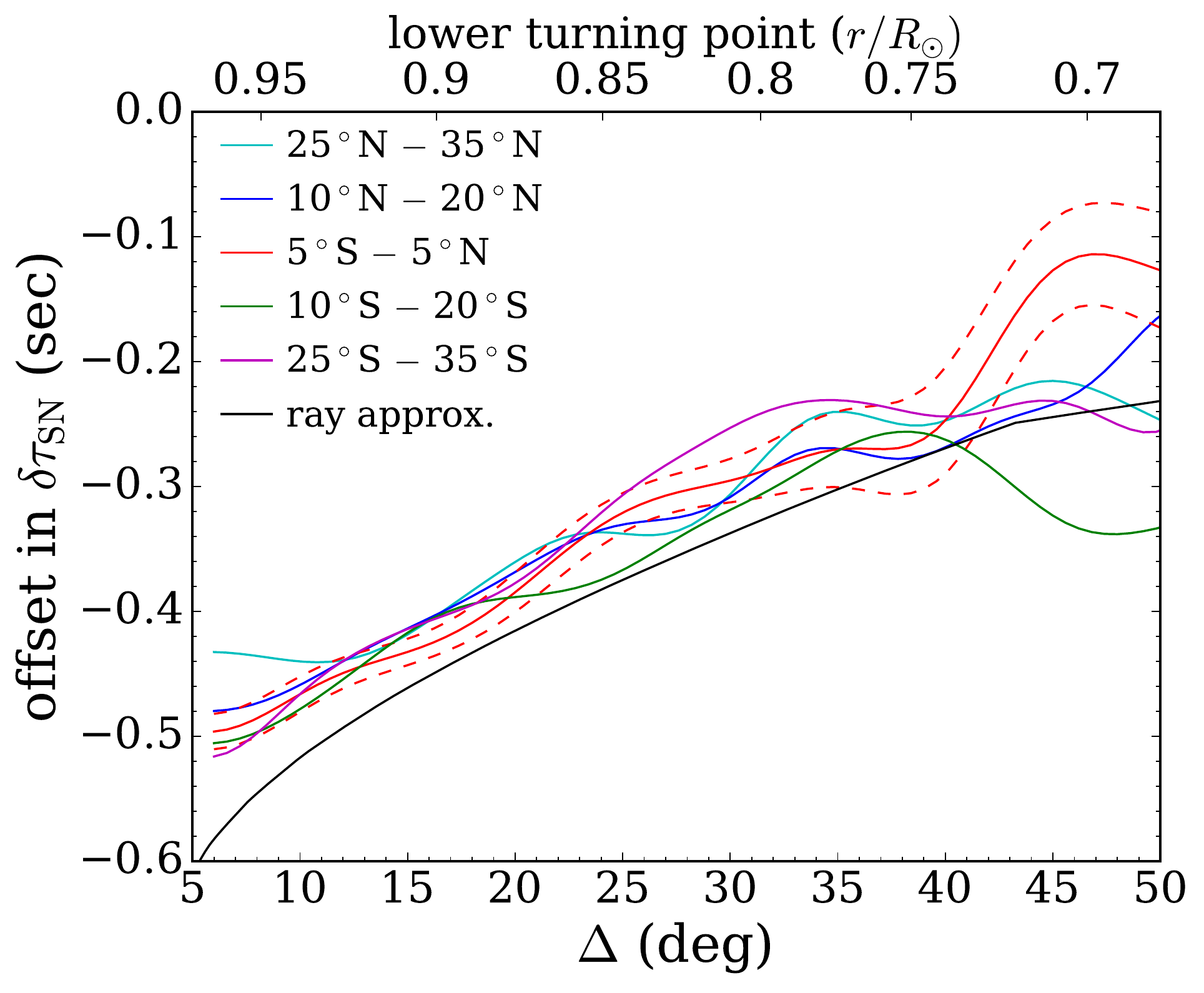}}
\caption{ \label{fig:offset}
Offsets between $P$-angle corrected and uncorrected $\delta\tau_\mathrm{SN}$ measured from MDI data as a function of travel distance $\Delta$.
A negative value corresponds to a southward flow owing to the $P$-angle error.
The offsets in $\delta\tau_\mathrm{SN}$ are averaged over 10$\degr$ bands around selected latitudes as labeled in the legend.
Gaussian smoothing, with FWHM=7.2$\degr$ in $\Delta$, is applied here.
The errors of only one curve, computed from the standard errors of the mean within each 7.2$\degr$ segment, are shown with dashed lines and those of other curves are similar.
The black solid line is a ray approximation of travel-time differences arising from a simple model of the leaking flow described in the text.
The corresponding radii of lower turning points from the ray approximation are indicated at the top.
}
\end{figure}

The $P$-angle error of MDI data gives rise to an offset in the $\delta\tau_\mathrm{SN}$ from MDI.
In order to investigate its influence, we compute the $\delta\tau_\mathrm{SN}$ from MDI without correcting the $P$-angle as well.
The travel-time offsets are obtained by subtracting the $P$-angle corrected $\delta\tau_\mathrm{SN}$ from the uncorrected $\delta\tau_\mathrm{SN}$.
Fig.~\ref{fig:offset} shows the offsets for selected latitudes as a function of travel distance.
The negative values correspond to a systematic southward flow presumably leaked from the rotation signal because of an additional counter-clockwise rotation of the camera with respect to the fixed Sun.
Evidently, the magnitude of these offsets decreases with travel distance.
Since the equatorial angular velocity in the convection zone is roughly independent of radius, the equator-crossing flow induced by the $P$-angle error should be proportional to the distance from the center of the Sun \citep[][, \S~6.2.3]{Giles2000} and hence weaker in the deeper layers.
To illustrate the point, we assume a constant differential rotation rate $\Omega=450$~nHz in the convection zone and a simple model of the leaking flow velocity
\begin{equation}
\vec{u} = \Omega\,r\sin\theta\,\sin(\delta P)\,\vec{\hat{\theta}},
\end{equation}
where $r$ is the distance from the Sun's center, \vec{\hat{\theta}} is the unit vector in the direction of increasing polar angle $\theta$, and $\delta P$ is set to be 0.21$\degr$.
The travel-time difference arising from this leaking flow in the ray approximation \citep{Kosovichev1997a} is estimated as
\begin{equation}
\delta\tau_\mathrm{SN} = -2\int_\Gamma\frac{1}{c^2} \, \vec{u}\cdot\mathrm{d}\vec{l},
\end{equation}
where the line integral is calculated along the ray path $\Gamma$ in the south-north direction across the equator, and $c$ is the sound speed from the solar model S \citep{Christensen-Dalsgaard1996}.
The result of the ray approximation shown in Fig.~\ref{fig:offset} is qualitatively in agreement with observations.
This result demonstrates a feasible way to discern whether a cross-equatorial signal is caused by a real south-north flow or merely an error in $P$-angle.
In the lower convection zone, these curves deviate from each other and have greater errors.
This indicates that without a proper treatment of the $P$-angle problem, a systematic error of the same order of magnitude as the meridional flow may be introduced for the measurement of meridional circulation even in the lower convection zone.
Similarly, we examine the offsets in $\delta\tau_\mathrm{EW}$ but find that the $P$-angle correction has almost no effect on $\delta\tau_\mathrm{EW}$ since a roll angle of 0.21$\degr$ only results in a 0.37\% leakage from the meridional flow ($\sim$0.04~m~s$^{-1}$) and a reduction of the tangential speed of the solar rotation by a factor of $7\times10^{-6}$ ($\sim$0.01~m~s$^{-1}$).

\begin{figure}
\centering
\resizebox{\hsize}{!}{\includegraphics{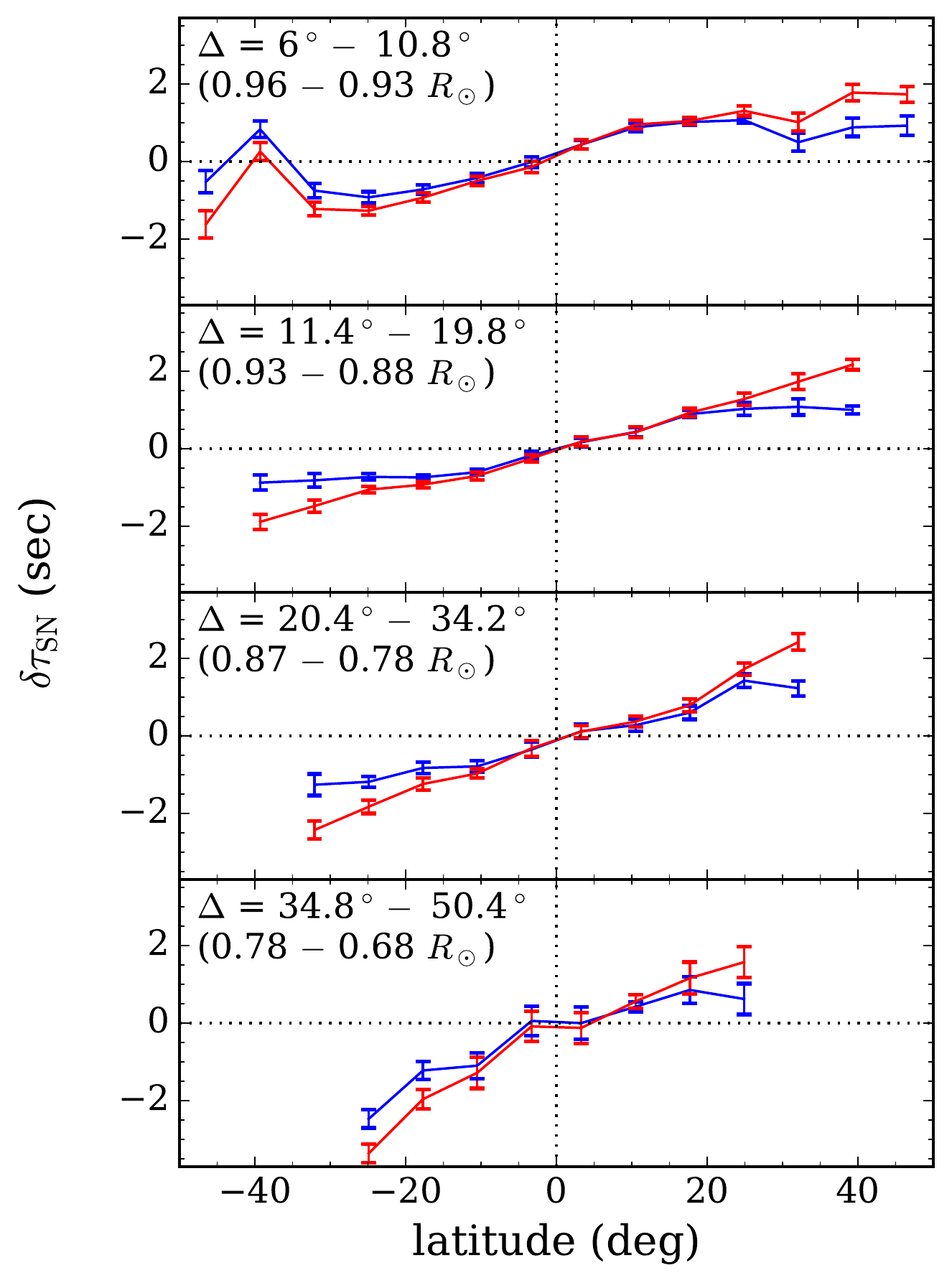}}
\caption{ \label{fig:cf-ns}
Comparisons of $\delta\tau_\mathrm{SN}$ between MDI (blue) and HMI (red) without center-to-limb correction as a function of latitude for different ranges of travel distances.
A $P$-angle correction of 0.21$\degr$ is applied to the MDI observations.
The range of travel distances and their corresponding radii of the lower turning point from the ray approximation are indicated in each panel; the shallowest is shown at the top and the deepest at the bottom.
The data values are binned every 7.2$\degr$ in latitude.
The error bars give the standard error of the mean in each binning interval.
}
\end{figure}

The comparison of $\delta\tau_\mathrm{SN}$ between MDI and HMI after MDI's $P$-angle has been corrected is shown in Fig.~\ref{fig:cf-ns}.
The antisymmetric part of the east-west travel-time differences, $\langle\delta\tau_\mathrm{EW}\rangle_\mathrm{antisym}$, representing the center-to-limb variation of MDI and HMI, are plotted in Fig.~\ref{fig:cf-we,ant}.
These $\delta\tau$s are averaged over four different ranges of travel distances and binned with an interval of 7.2$\degr$ in latitude (or longitude).
Obviously, the $\delta\tau_\mathrm{SN}$s are dominated by the center-to-limb variation throughout the entire convection zone since they both have the same order of magnitude.
Also, they grow with travel distance and become so large for meridional circulation at the base of convection zone that is at odds with our current theoretical picture of the Sun \citep{Duvall2009}.
A similar phenomenon is, unsurprisingly, found in GONG data \citep{Kholikov2014}.
In addition, the large-scale center-to-limb variation in MDI and HMI are quite unlike each other.
The discrepancies between the two data sets are several times larger than the error bars at mid- and high latitudes.
Despite the disagreement at large scales, their small-scale variations are fairly consistent.

\begin{figure}
\centering
\resizebox{\hsize}{!}{\includegraphics{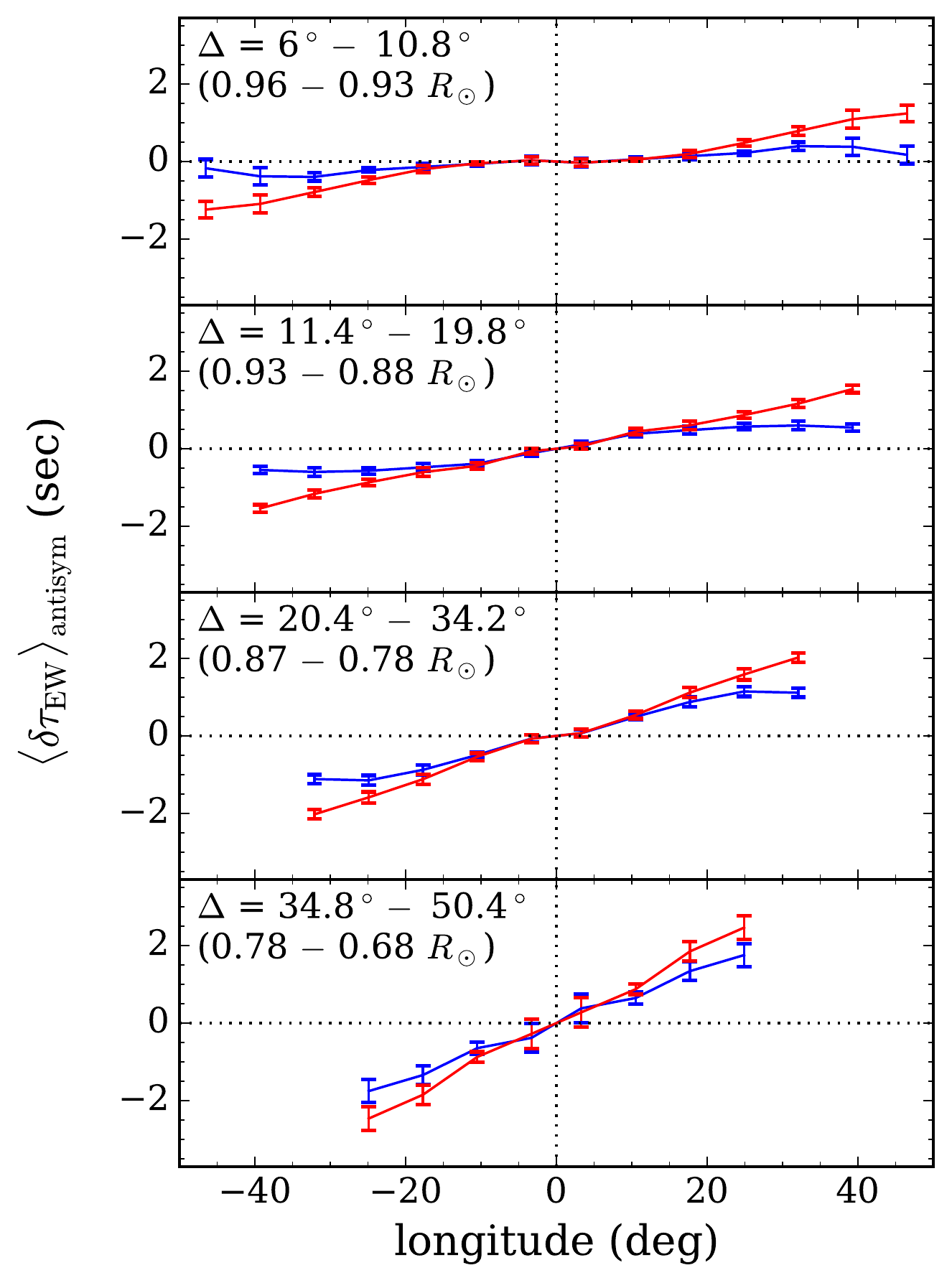}}
\caption{ \label{fig:cf-we,ant}
Same as Fig.~\ref{fig:cf-ns}, but the travel-time differences are measured in the east-west direction and antisymmetrized about the central meridian.
}
\end{figure}

To explore the systematic errors further, the symmetric part of the east-west travel-time differences $\langle\delta\tau_\mathrm{EW}\rangle_\mathrm{sym}$ is plotted for both data sets in Fig.~\ref{fig:cf-we,sym}.
Since the Dopplergrams have been tracked with the differential rotation rate at the surface, the overall nonzero signals are caused by the difference between the surface and internal solar rotation and are expected to be constant values along the equator.
However, the $\langle\delta\tau_\mathrm{EW}\rangle_\mathrm{sym}$ from MDI and HMI are neither constant with longitude nor in agreement, thus revealing another systematic variation.
This systematic effect only appears in the east-west direction, as the $\delta\tau_\mathrm{SN}$ does not have a symmetric component on the order of one second.
Although the systematic error in $\langle\delta\tau_\mathrm{EW}\rangle_\mathrm{sym}$ does not affect the meridional flow measurement, it has an impact on the time-distance measurement of the solar rotation.

\begin{figure}
\centering
\resizebox{\hsize}{!}{\includegraphics{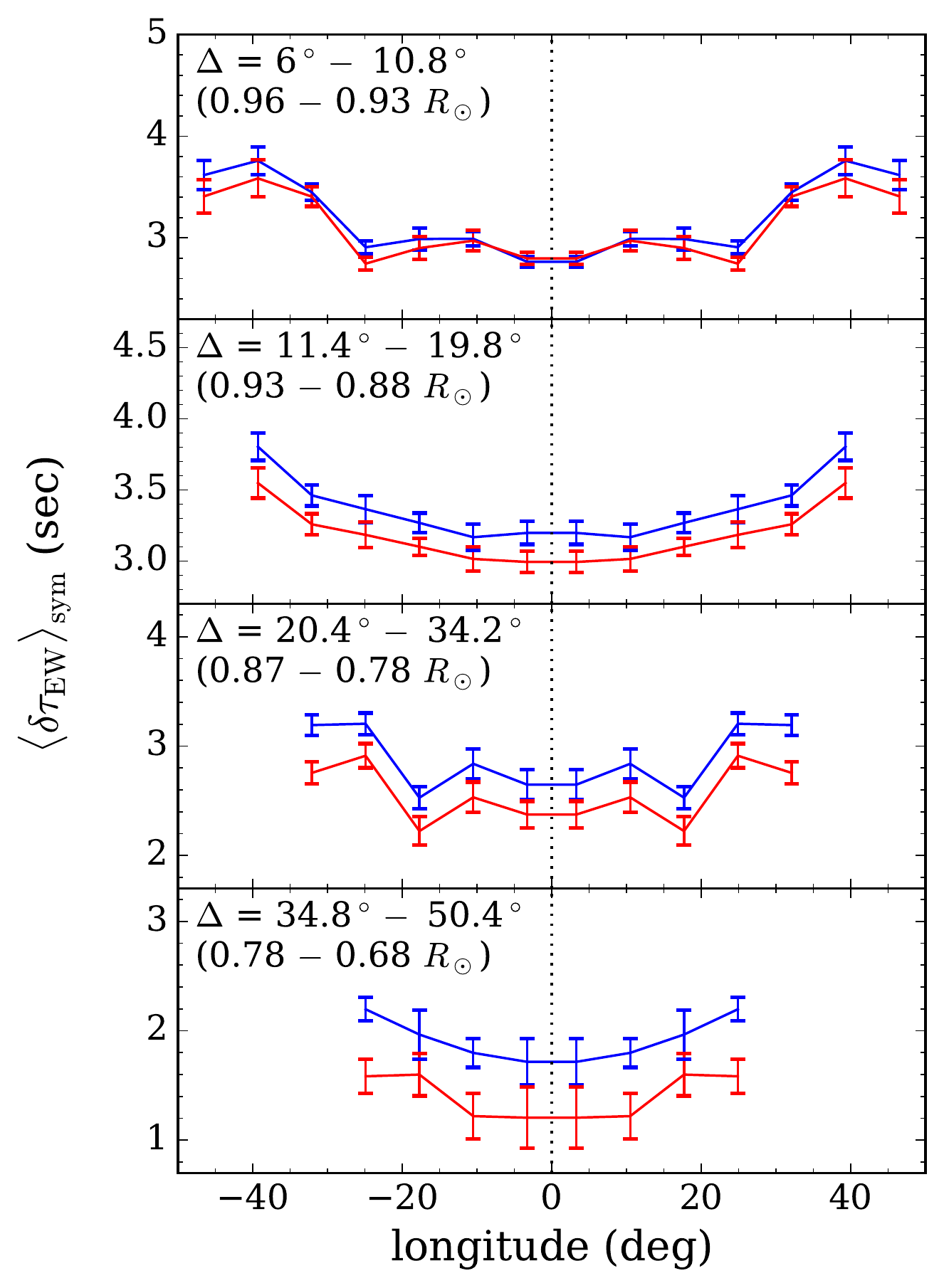}}
\caption{ \label{fig:cf-we,sym}
Same as Fig.~\ref{fig:cf-we,ant}, but the symmetric part of the east-west travel-time differences are shown instead of the antisymmetric part.
}
\end{figure}

\begin{figure}
\centering
\resizebox{\hsize}{!}{\includegraphics{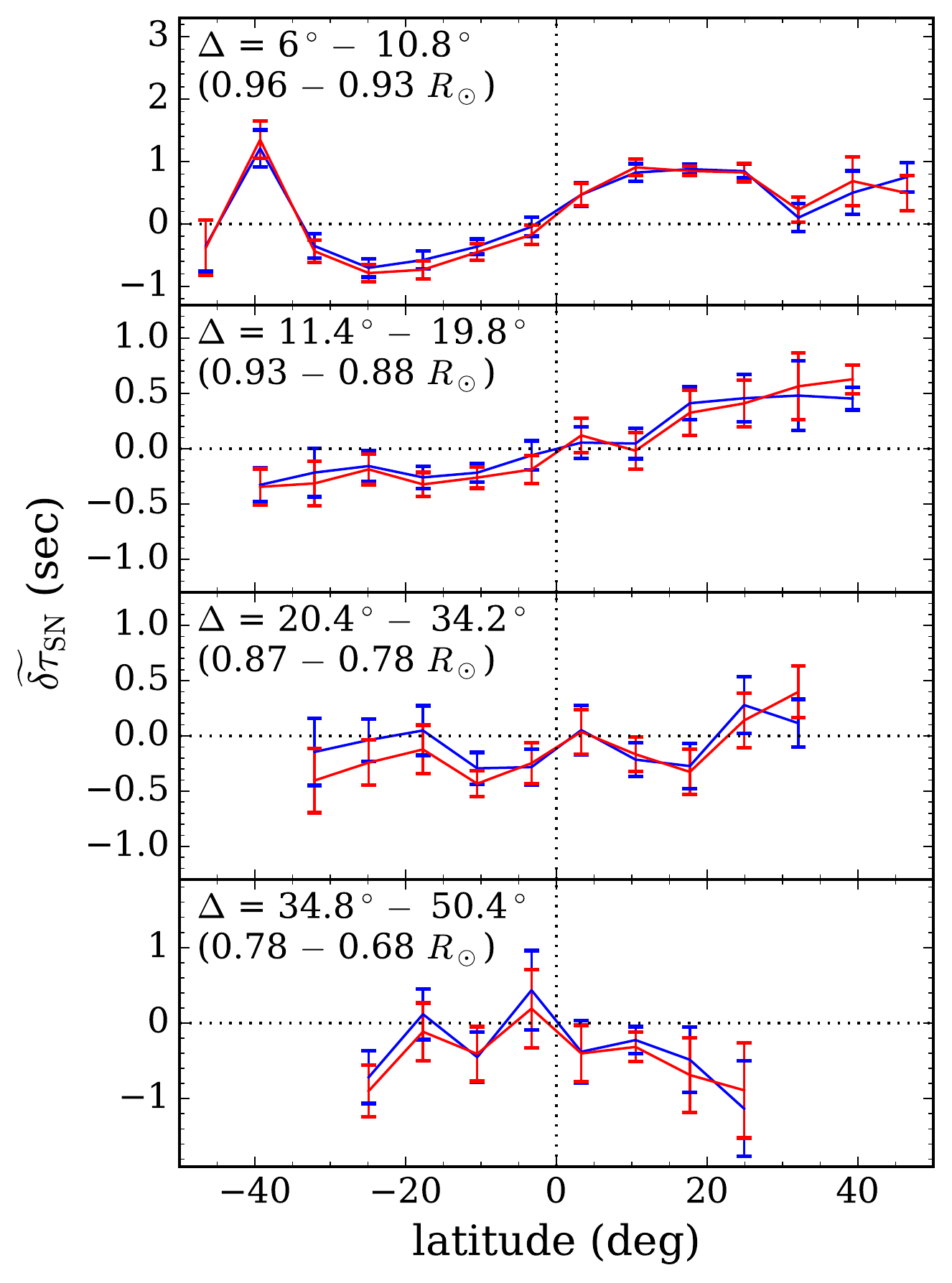}}
\caption{ \label{fig:cf-sm}
Same as in Fig.~\ref{fig:cf-ns} but the center-to-limb variation of both data sets have been removed by subtracting $\langle\delta\tau_\mathrm{EW}\rangle_\mathrm{antisym}$ from $\delta\tau_\mathrm{SN}$; the vertical scales are different from Fig.~\ref{fig:cf-ns}.
}
\end{figure}

The removal of the center-to-limb variation is carried out by subtracting $\langle\delta\tau_\mathrm{EW}\rangle_\mathrm{antisym}$ from $\delta\tau_\mathrm{SN}$ as suggested by \citet{Zhao2012}.
The comparison of center-to-limb corrected $\widetilde{\delta\tau}_\mathrm{SN}$ between MDI and HMI is shown in Fig.~\ref{fig:cf-sm}.
This ad~hoc correction reduces the discrepancies between the two data sets to within one standard error in all travel-distance ranges, and significantly altered the magnitudes and even the signs of $\delta\tau_\mathrm{SN}$.
We note that $\langle\delta\tau_\mathrm{EW}\rangle_\mathrm{antisym}$ is much greater than the desired signal $\widetilde{\delta\tau}_\mathrm{SN}$, especially for the measurements of flows in the lower convection zone.
Also the error bars in the lower panels of Fig.~\ref{fig:cf-sm} are as large as the signals after averaging over 288 days data and a large range of travel distances.
Both the considerable magnitude of the center-to-limb variation and the low signal-to-noise ratio for measurements at large distances demonstrate the difficulty of obtaining an accurate travel-time measurement of deep meridional flow.
The error bars at high latitudes in the top panel of Fig.~\ref{fig:cf-sm} are rather large because of the effects of foreshortening in combination with the spatial sampling \citep{Beck2005}: for short distance cases, both of the paired points used to compute the CCF are close to the high-latitude areas and the number of points on the arc is small as well.
We also note a small nonzero signal at equator, whose value is well within the error bar though.
This nonzero $\widetilde{\delta\tau}_\mathrm{SN}$ at the equator is unlikely to be caused by an instrumental pointing issue because the trend is not the same as that in Fig.~\ref{fig:offset}.
It would be of interest if this nonzero signal still exists after averaging additional data since the cross-equatorial flux transport plays a role in some solar dynamo models \citep{Cameron2013}.

\begin{figure}
\centering
\resizebox{\hsize}{!}{\includegraphics{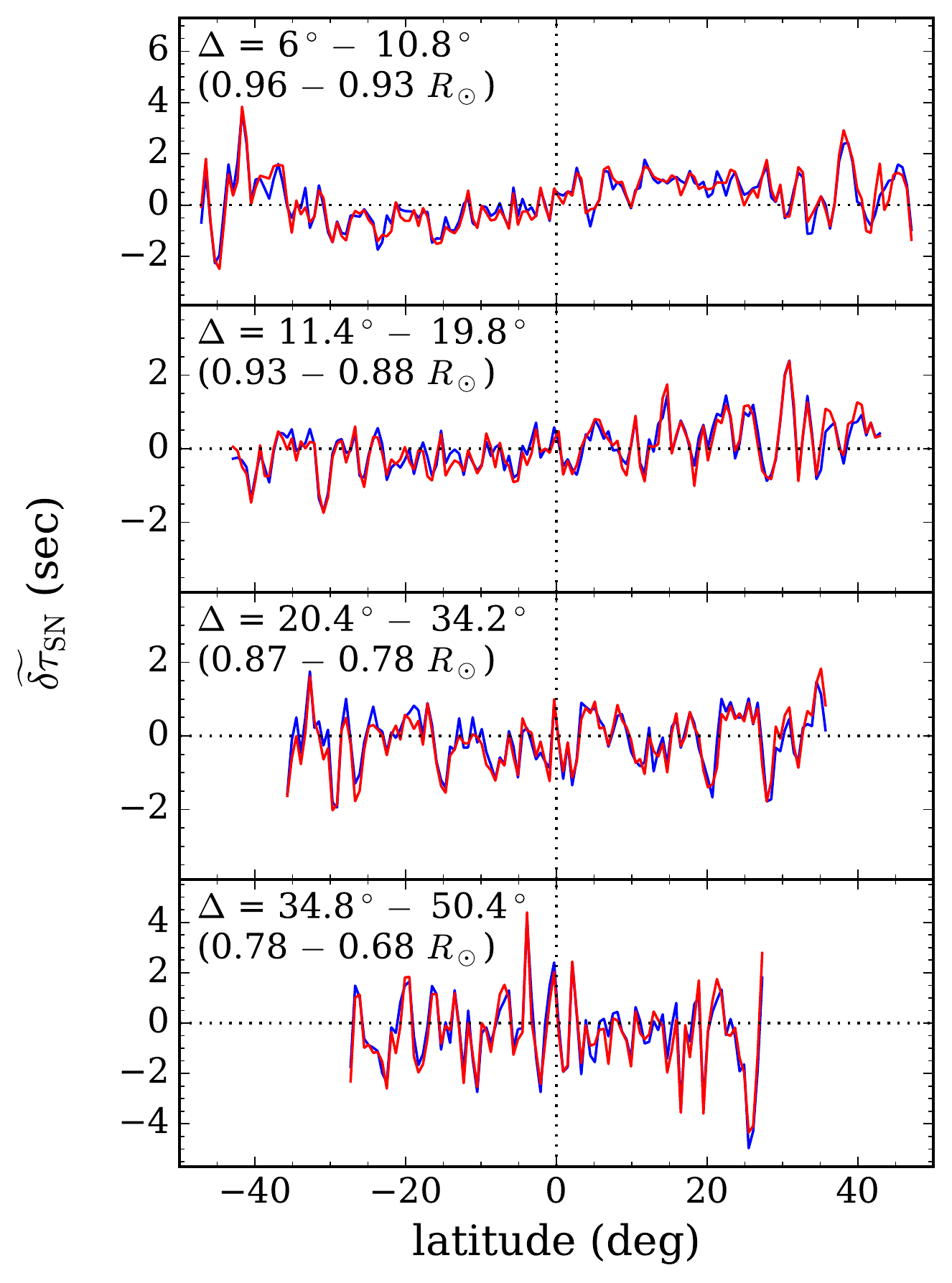}}
\caption{ \label{fig:cf-orig}
Same as in Fig.~\ref{fig:cf-sm}, but for $\widetilde{\delta\tau}_\mathrm{SN}$ without binning in latitude.
The small-scale fluctuations of both data sets match each other very well, indicating their solar origin.
}
\end{figure}

\begin{figure}
\centering
\resizebox{\hsize}{!}{\includegraphics{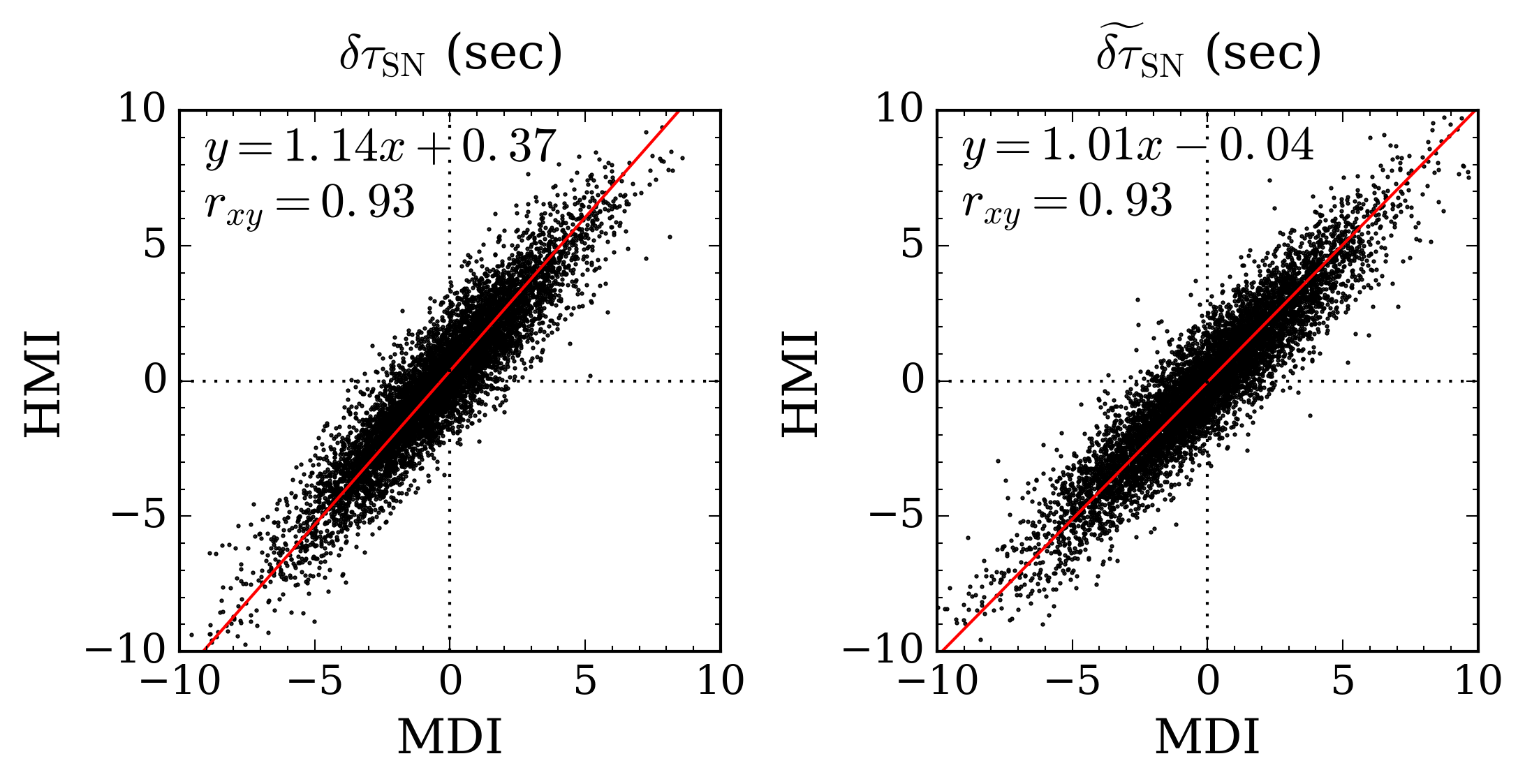}}
\caption{ \label{fig:rxy}
\emph{Left}: Scatter plot of $\delta\tau_\mathrm{SN}$ from HMI ($y$-axis) vs. MDI ($x$-axis) for all available latitudes and travel distances ranging from 10.2$\degr$ to 45$\degr$.
The center-to-limb variation and MDI $P$-angle error are not corrected.
\emph{Right}: Same as in the left panel but with corrections to the center-to-limb variation and the MDI $P$-angle.
The red lines refer to linear functions that fit the data points assuming equal errors in both coordinates \citep[][, \S~15.3]{Press1992}.
The Pearson correlation coefficients, $r_{xy}$, between the two data sets and the fitted linear functions are labeled in each panel.
}
\end{figure}

To scrutinize the consistency between MDI and HMI at small scales, the $\widetilde{\delta\tau}_\mathrm{SN}$ without binning is shown in Fig.~\ref{fig:cf-orig}.
Noticeably, the small-scale fluctuations of $\widetilde{\delta\tau}_\mathrm{SN}$ from the two data sets match each other closely for all travel-distance ranges.
For a quantitative comparison of the fluctuations, the scatter plots of travel-time differences from MDI versus HMI are shown in Fig.~\ref{fig:rxy}.
Both the corrected and uncorrected travel-time differences are plotted for comparison.
The slopes and offsets of the linear fits to the scatter plots represent the large-scale systematics of the center-to-limb variation and $P$-angle error respectively.
The $P$-angle correction to MDI images removes the 0.37-sec offset while the center-to-limb correction adjusts the slope to unity, that is to say, MDI and HMI results are consistent at large scales after the two major corrections.
On the other hand, the coherence of small-scale fluctuations is indicated by high correlations (0.93) between MDI and HMI, which suggests that these fluctuations are of solar origin since they are present in both data sets.
Unlike the significant influence of corrections on large scale, the effect of the corrections on the Pearson correlation coefficients is inconsequential.
In this work we do not smooth the $\delta\tau_\mathrm{EW}$ when applying center-to-limb correction.
In other words, the $\widetilde{\delta\tau}_\mathrm{SN}$ consists of the fluctuations both from $\delta\tau_\mathrm{SN}$ and $\delta\tau_\mathrm{EW}$.
The persistence of the correlation after the center-to-limb correction implies that the degree of consistency of $\delta\tau_\mathrm{EW}$ in small scale is as high as that of $\delta\tau_\mathrm{SN}$.

\section{Summary and discussion}
\label{sec:discuss}

By comparing the time-distance measurements from MDI data with those from HMI data, we explore the influences of two major systematic errors, the MDI $P$-angle misalignment and the center-to-limb variation, on the acoustic travel-time differences arising from meridional circulation.
The methods for correcting the systematics are examined, assuming that contemporaneous data from different instruments should give the same meridional flow measurement in spite of different observation heights in the solar atmosphere.

We determine the $P$-angle drift $\delta P$ of MDI by cross-correlating medium-$\ell$ Dopplergrams from MDI with those from HMI (front camera) and found a temporal variation.
Some of the discontinuities in the time-varying $\delta P$ are related to the changes of the SOHO spacecraft flight operation.
The $\delta P$ leads to a travel-distance-dependent offset in the south-north travel-time differences $\delta\tau_\mathrm{SN}$ from MDI.
This travel-time offset decreases with increasing travel distance.
Also, the offsets at different latitudes are in line with each other in this regime.
However, these offsets begin to fluctuate and depart from each other as the lower turning points to which the travel distances correspond reach the lower convection zone.
This indicates that a $P$-angle problem should be handled with caution for the deep meridional flow measurements, otherwise it would introduce a systematic error of the same order of magnitude as the meridional flow in the convection zone.

The $\delta\tau_\mathrm{SN}$ measured from MDI and HMI data are dominated by large-scale center-to-limb variations, which grow with travel distance and have different patterns for MDI and HMI.
The antisymmetrized east-west travel-time differences $\langle\delta\tau_\mathrm{EW}\rangle_\mathrm{antisym}$ represent the center-to-limb variation to a large extent.
After subtracting $\langle\delta\tau_\mathrm{EW}\rangle_\mathrm{antisym}$ from $\delta\tau_\mathrm{SN}$, the center-to-limb corrected $\widetilde{\delta\tau}_\mathrm{SN}$ from MDI and HMI agree with each other within the 288-day error bars.
In \citet{Zhao2012}, they only examined the center-to-limb correction at a near-surface layer with 10 days of data.
Our results show that this ad~hoc method seems to be valid throughout the entire convection zone, even at the base, which gives us the confidence in the correction and measurement of the deep meridional flow.
In addition, the small-scale fluctuations in both data sets are highly correlated, which implies that they are of solar origin.
The highly correlated fluctuations between MDI and HMI data are also observed in the measurements of supergranulation \citep{Svanda2013,Williams2014} and in a direct comparison of Dopplergrams as shown in Sect.~\ref{roll} and in \citet{Howe2011a}.
Most of these fluctuations should be ascribed to realization noise due to the stochastic excitation of acoustic waves by turbulent convection, but they may also include scattering by intermediate-scale convection \citep{Gizon2004}.

The shallowest measurements of $\widetilde{\delta\tau}_\mathrm{SN}$, shown in the top panel of Fig.~\ref{fig:cf-sm}, exhibit some irregular features at high latitudes.
One likely cause is that the medium-$\ell$ data have little $p$-mode power for waves traveling in the near-surface layers.
In addition, the foreshortening effect reduces the image resolution at high latitudes as mentioned in Sect.~\ref{sec:result}.
As a consequence, the $\widetilde{\delta\tau}_\mathrm{SN}$, merely averaged over 288 days in this work, are severely influenced by realization noise at high latitudes for the shallow measurements.
We also note that a rapid change in $\delta\tau$ at high latitudes appears in the GONG data analysis by \citet{Kholikov2014}.
They suggested that the possible causes for the variations at high latitudes are either a systematic effect resulting from the solar $B_0$ variation or the use of the center-to-limb correction method.

\begin{figure}
\centering
\resizebox{\hsize}{!}{\includegraphics{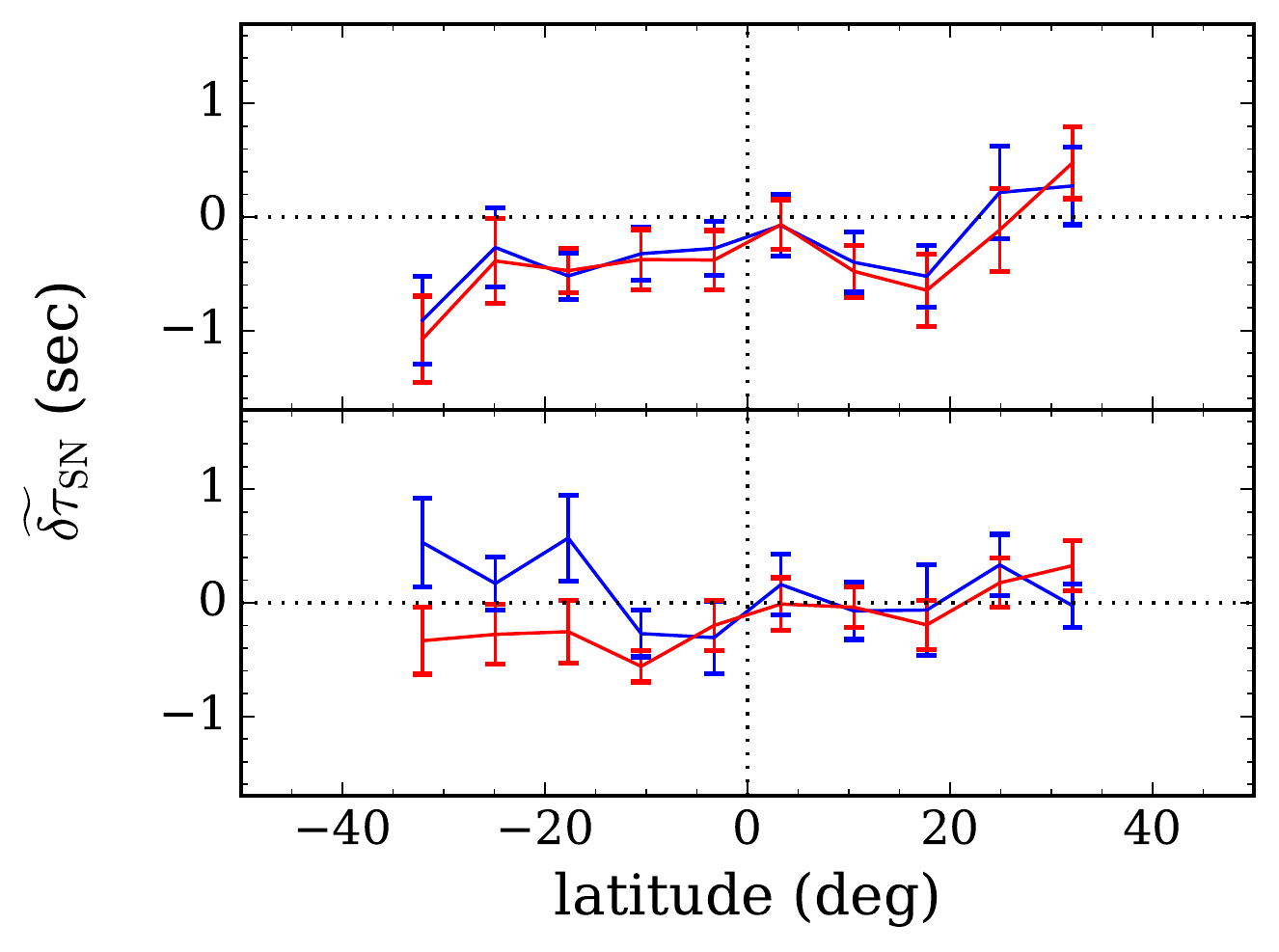}}
\caption{ \label{fig:flip}
Comparison of $\widetilde{\delta\tau}_\mathrm{SN}$ between MDI (blue) and HMI (red) using data in the periods when SOHO spacecraft was ``upright'' (upper) and ``upside down'' (lower).
The data are averaged over the travel-distance range 20.4$\degr$--34.2$\degr$ as in the third panel of Fig.~\ref{fig:cf-sm}.
Unlike in the non-flipping periods, the discrepancies between the two curves become noticeable in the flipping periods.
}
\end{figure}

In the course of data analysis we find that the MDI $\delta\tau$ give inconsistent results when the SOHO spacecraft was ``upside down'', which had been reported in the past \citep{Duvall2009,Howe2011,Liang2015a,Liang2015b}.
Fig.~\ref{fig:flip} shows the $\widetilde{\delta\tau}_\mathrm{SN}$ from MDI and HMI averaging over the periods when SOHO was flipped ($\mathtt{CROTA2} > 170\degr$; 151 days) and upright ($\mathtt{CROTA2} < 10\degr$; 137 days), respectively.
Comparing Fig.~\ref{fig:flip} to the third panel in Fig.~\ref{fig:cf-sm}, it appears that the discrepancies of $\widetilde{\delta\tau}_\mathrm{SN}$ between the two data sets largely come from the MDI flipped periods.
While the calibration of MDI image distortion does improve the correlation between MDI and HMI, the inconsistency due to the instrument flipping remains unresolved.
This systematic error would become conspicuous when multiple-year data are utilized in which the noise level would be much lower than this work.
Besides, it is inconvenient that MDI instrument always flipped in the periods when the solar tilt angle $B_0$ is close to zero, which places one in a dilemma of choosing between large-$B_0$ images or images for which MDI is flipped when studying a long-term variation of meridional circulation with time-distance analysis.
It would be useful in future work to develop a better model of MDI image distortion and of MDI point-spread function (PSF), which may help understand the physical origin of this systematic error.
In particular, the imperfect knowledge of the MDI PSF has been shown to be a limitation for high-degree modes \citep{Korzennik2004,Korzennik2013}.
It might also be useful to better understand the Doppler calibration and height sensitivity for the two instruments.

Although our study shows a remarkable degree of consistency between MDI and HMI, some limitations and caveats in this work have to be addressed.
First, this comparison cannot rule out the systematic errors affecting both data sets in the same way.
Second, it had been estimated that a reliable measurement for the meridional circulation at the base of the convection zone requires more than one decade of continuous observations \citep{Braun2008}.
Thus, the signal-to-noise ratio of our 288-day measurement is far from enough to show that the ad~hoc correction suffices for the deep meridional flow measurement.
Finally, in this article we have not studied time-varying systematic errors, such as the error related to the annual variation of solar $B_0$ angle \citep{GonzalezHernandez2006,Zaatri2006} and the long-term variation of the center-to-limb effect \citep{Liang2015b}, which would require longer contemporaneous data sets.
Hence, it would be worth comparing more than a decade of GONG data with the MDI or HMI data in future work.

\begin{acknowledgements}
The HMI data used are courtesy of NASA/SDO and the HMI science team.
SOHO is a project of international cooperation between ESA and NASA.
The data were processed at the German Data Center for SDO (GDC-SDO), funded by the German Aerospace Center (DLR).
Support is acknowledged from the SpaceInn and SOLARNET projects of the European Union.
L.G. acknowledges support from the NYU Abu Dhabi Center for Space Science under grant no. G1502.
We used the workflow management system Pegasus funded by The National Science Foundation under OCI SI2-SSI program grant \#1148515 and the OCI SDCI program grant \#0722019.
\end{acknowledgements}

\bibliography{}

\begin{appendix}

\section{Position of the fitting window}
\label{suppl:win}

\begin{figure}[h]
\centering
\resizebox{\hsize}{!}{\includegraphics{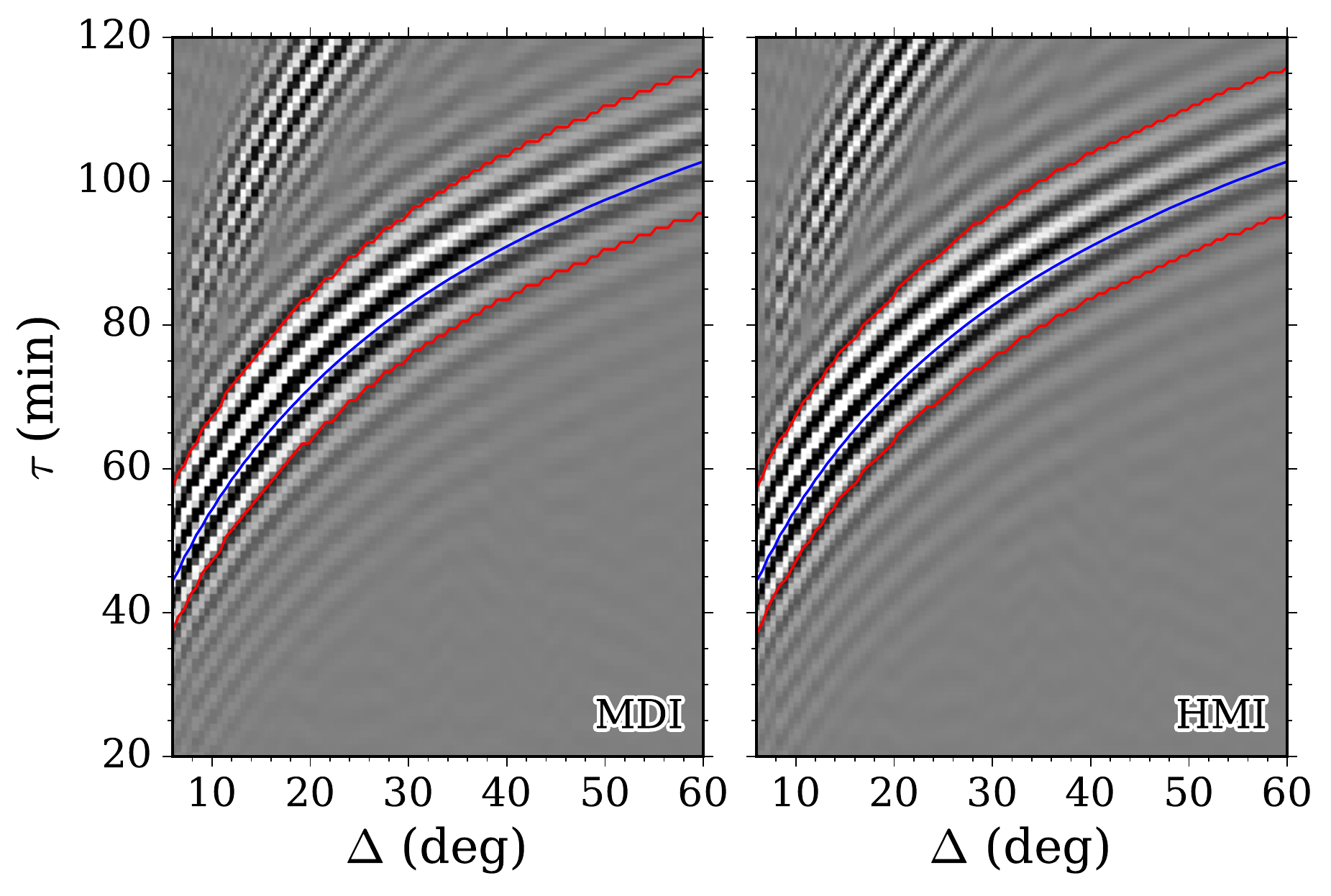}}
\caption{ \label{fig:ccf}
Examples of 288-day averaged cross-correlation function obtained from MDI (left) and HMI (right) as a function of distance, $\Delta$, and time lag, $\tau$.
They are computed in the south-north direction and averaged over $\pm15\degr$~longitude and $\pm15\degr$~latitude.
A 20-minute window that isolates the first-skip branch of the cross-correlation function is used in the fitting (20 data points for MDI and 27 data points for HMI).
In both panels, the blue curve represents the fitted phase time and the red curves indicate the border of the fitting window for each $\Delta$.
The center positions of the fitting window and the fitted phase times for every 3$\degr$ of travel distance are listed in Table~\ref{tab:win}.
}
\end{figure}

\begin{table}[h]
\caption{Center positions of the fitting window and the fitted phase times} 
\label{tab}
\centering
\begin{tabular}{c c c c c c}
\hline\hline 
$\Delta$ & $\tau_\mathrm{g}^\mathrm{(ray)}$ & $\tau_\mathrm{ctr}^\mathrm{(MDI)}$ & $\tau_\mathrm{ctr}^\mathrm{(HMI)}$ & $\tau_\mathrm{ph}^\mathrm{(MDI)}$ & $\tau_\mathrm{ph}^\mathrm{(HMI)}$ \\
(deg) & (min) & (min) & (min) & (min) & (min) \\
\hline 
\phantom{0}6 & \phantom{0}44.79 & \phantom{0}47.5 & \phantom{0}47.25 & \phantom{0}44.44 & \phantom{0}44.42 \\
\phantom{0}9 & \phantom{0}52.55 & \phantom{0}55.5 & \phantom{0}54.75 & \phantom{0}52.02 & \phantom{0}52.02 \\
          12 & \phantom{0}58.84 & \phantom{0}61.5 & \phantom{0}61.50 & \phantom{0}58.49 & \phantom{0}58.49 \\
          15 & \phantom{0}64.17 & \phantom{0}66.5 & \phantom{0}66.75 & \phantom{0}63.81 & \phantom{0}63.80 \\
          18 & \phantom{0}68.82 & \phantom{0}71.5 & \phantom{0}71.25 & \phantom{0}68.50 & \phantom{0}68.49 \\
          21 & \phantom{0}72.94 & \phantom{0}75.5 & \phantom{0}75.75 & \phantom{0}72.61 & \phantom{0}72.61 \\
          24 & \phantom{0}76.62 & \phantom{0}79.5 & \phantom{0}78.75 & \phantom{0}76.31 & \phantom{0}76.31 \\
          27 & \phantom{0}79.95 & \phantom{0}82.5 & \phantom{0}82.50 & \phantom{0}79.65 & \phantom{0}79.64 \\
          30 & \phantom{0}82.97 & \phantom{0}85.5 & \phantom{0}85.50 & \phantom{0}82.66 & \phantom{0}82.66 \\
          33 & \phantom{0}85.71 & \phantom{0}88.5 & \phantom{0}88.50 & \phantom{0}85.40 & \phantom{0}85.39 \\
          36 & \phantom{0}88.23 & \phantom{0}90.5 & \phantom{0}90.75 & \phantom{0}87.89 & \phantom{0}87.88 \\
          39 & \phantom{0}90.53 & \phantom{0}93.5 & \phantom{0}93.00 & \phantom{0}90.18 & \phantom{0}90.18 \\
          42 & \phantom{0}92.63 & \phantom{0}95.5 & \phantom{0}95.25 & \phantom{0}92.32 & \phantom{0}92.31 \\
          45 & \phantom{0}94.58 & \phantom{0}97.5 & \phantom{0}96.75 & \phantom{0}94.29 & \phantom{0}94.28 \\
          48 & \phantom{0}96.43 & \phantom{0}98.5 & \phantom{0}99.00 & \phantom{0}96.21 & \phantom{0}96.21 \\
          51 & \phantom{0}98.20 &           100.5 &           100.50 & \phantom{0}97.94 & \phantom{0}97.93 \\
          54 & \phantom{0}99.90 &           102.5 &           102.75 & \phantom{0}99.63 & \phantom{0}99.63 \\
          57 &           101.53 &           104.5 &           104.25 &           101.21 &           101.19 \\
          60 &           103.08 &           105.5 &           105.75 &           102.74 &           102.73 \\
\hline 
\end{tabular}
\tablefoot{ \label{tab:win}
The group time, $\tau_\mathrm{g}^\mathrm{(ray)}$, based on the ray approximation is used as the reference time to determine the fitting window.
Modes with frequency of 3.33~mHz are used in the ray approximation.
Because the theoretical $\tau_\mathrm{g}^\mathrm{(ray)}$ are roughly half a period (2.5~minutes) smaller than the observed group times (the center position of the fitted Gaussian envelope) for travel distances in this regime, we round off the $\tau_\mathrm{g}^\mathrm{(ray)}$ plus 2.5~minutes to the nearest grid point as the center position of the fitting window, $\tau_\mathrm{ctr}$.
The position of CCF's peak closest to $\tau_\mathrm{g}^\mathrm{(ray)}$ is used as the initial guess for the phase time, $\tau_\mathrm{ph}$, in the fitting.
The fitted $\tau_\mathrm{ph}$ for the CCFs in Fig.~\ref{fig:ccf} are listed in the table for comparison.
}
\end{table}

\end{appendix}

\end{document}